\documentclass[12pt]{article}
\usepackage{amssymb}
\usepackage{graphicx}
\newlength{\defbaselineskip}
\newcommand{\setlinespacing}[1]%
           {\setlength{\baselineskip}{#1 \defbaselineskip}}

\newcommand{\be}{\begin{equation}}
\newcommand{\ee}{\end{equation}}
\newcommand{\bea}{\begin{eqnarray}}
\newcommand{\eea}{\end{eqnarray}}
\newcommand{\bd}{\begin{displaymath}}
\newcommand{\ed}{\end{displaymath}}
\newcommand{\bas}{\begin{eqnarray*}}
\newcommand{\eas}{\end{eqnarray*}}
\newcommand{\ba}{\begin{array}}
\newcommand{\ea}{\end{array}}
\newcommand{\bq}{\begin{quote}}
\newcommand{\eq}{\end{quote}}
\newcommand{\bds}{\begin{description}}
\newcommand{\eds}{\end{description}}
\newcommand{\ben}{\begin{displaymath}}
\newcommand{\een}{\end{displaymath}}
\newcommand{\R}{\mbox{$\Bbb R$}}
\newcommand{\sech}{\mathop{\rm sech}\nolimits}
\newcommand{\csch}{\mathop{\rm csch}\nolimits}
\newcommand{\ca}{{\cal A}}
\newcommand{\bca}{\bar{\cal A}}

  \oddsidemargin
 0.2in \evensidemargin 0.6in
\title{
$\cal PT$-symmetric supersymmetry in a solvable short-range model
}
\author{B Bagchi$^a$, H B\'\i la$^b$, V Jakubsk\'y$^b$, S Mallik$^a$, C Quesne$^c$,
M Znojil$^b$\\
{\small $^a$ Department of Applied Mathematics, University of Calcutta,} \\
{\small 92 Acharya Prafulla Chandra Road, Kolkata 700 009, West Bengal, India}\\
{\small $^b$ \'Ustav jadern\'e fyziky AV \v CR, 250 68 \v Re\v z, Czech Republic}\\
{\small $^c$ Physique Nucl\'eaire Th\'eorique et Physique Math\'ematique,} \\
{\small Universit\'e Libre de Bruxelles, Campus de la Plaine CP229,} \\
{\small Boulevard~du
Triomphe, B-1050 Brussels, Belgium} \\
{\small E-mail: bbagchi123@rediffmail.com, Hynek.Bila@ujf.cas.cz,
jakub@ujf.cas.cz,
} \\
{\small smallik123@rediffmail.com, cquesne@ulb.ac.be and
znojil@ujf.cas.cz}}
\date{ }
\begin{document}
\baselineskip=22pt plus 1pt minus 1pt
%%%%%%%%%%%%%%%%%%%%%%%%%%%%%%%%%%%%%%%%%%%%%%%%%%%%%%%%%%
\maketitle

\begin{abstract}
The simplest purely imaginary and piecewise constant $\cal
PT$-symmetric potential located inside a larger box is studied.
Unless its strength exceeds a certain critical value, all the
spectrum of its bound states remains real and discrete. We
interpret such a model as an initial element of the generalized
non-Hermitian Witten's hierarchy of solvable Hamiltonians and
construct its first supersymmetric (SUSY) partner in closed form.

\end{abstract}


%
%========================================================================
%
\newpage

\section{Introduction and summary \label{sec:intro}}

% \subsection{Supersymmetry (SUSY) \label{sec:susy}}

One of the most intriguing {experimental} puzzles encountered in
contemporary physics is the evident absence of SUSY partners of
elementary particles in nature. In the context of {field theory}
this means that SUSY, if it exists, must be spontaneously broken.
Witten  \cite{Witten} proposed a schematic model which,
incidentally, failed to clarify this breakdown but, nonetheless,
survived and found a number of applications within the
so-called SUSY quantum mechanics (SUSYQM) \cite{CKS}.\par
%
%-----------------------------------------------------
%
In the latter formalism one introduces the so-called
superpotential $W(x)$ and defines the two operators
\be
  \ca = \frac{d}{dx} + W(x) \qquad \bca = - \frac{d}{dx} + W(x)
\ee
with the property that the two related {\em different} (so-called
`SUSY partner') potentials $ V^{(\pm)}- E_0 = W^2 \mp W'$ may
prove {\em both} exactly solvable at the same time. An easy
explanation of this phenomenon lies in the fact that the related
Hamiltonians
\be
  H^{(\pm)} = - \frac{d^2}{dx^2} + V^{(\pm)}(x) - E_0
  \label{eq:Hpm}
\ee
become inter-related, at a convenient auxiliary energy $E = E_0$,
by the factorization rules $H^{(+)}=\bca \ca$ and $H^{(-)}=\ca
\bca$. The spectra of $H^{(+)}$ and $H^{(-)}$ are then alike
except possibly for the ground state. In the unbroken SUSY case,
the ground state at vanishing energy is nondegenerate and, in the
present notational set-up, it belongs to $H^{(+)}$. This means
that
\be
  \ca\, \psi^{(+)}_0(x) = 0 \label{eq:unbroken-susy}
\ee
where $\psi^{(+)}_n(x)$ (resp.\ $\psi^{(-)}_n(x)$), $n=0$, 1,
2,~\ldots, denote the wavefunctions of $H^{(+)}$ (resp.\
$H^{(-)}$). The (double) degeneracy of $\left(\psi^{(+)}_{n+1}(x),
\psi^{(-)}_n(x)\right)$ for $n=0$, 1, 2,~\ldots\ is implied by the
intertwining relationships
\be
  \ca\, H^{(+)} = H^{(-)} \ca \qquad H^{(+)} \bca
  = \bca\, H^{(-)}. \label{eq:intertwining}
\ee
In the conventional setting, the Hamiltonians (\ref{eq:Hpm}) are
assumed self-adjoint.\par
%
%--------------------------------------------------------
%
New horizons have been opened by the pioneering letter by Bender
and Boettcher \cite{BB} who noticed, in a slightly different
context, that the latter condition $H = H^\dagger$ might be
relaxed as redundant and replaced by its suitable weakened forms.
For our present purposes, we shall employ their proposal and in equation
(\ref{eq:Hpm}) allow complex potentials that are merely constrained by the
requirement that their real and imaginary parts are spatially symmetric
and antisymmetric, respectively \cite{BBjmp}.\par
%
%-------------------------------------------------------
%
It is not too difficult to show that the above SUSYQM
factorization scheme remains unchanged under such a non-Hermitian
generalization~\cite{Andrianov,crea,ptsusy,BMQ1,CQ}. Of course,
the relaxation of the usual condition $H=H^\dagger$ is by far not
a trivial step. Formally, we may put $H^\dagger = {\cal T}H{\cal
T}$ with an antilinear `time-reversal' operator ${\cal T}$
\cite{erratum}. In such a setting, Bender and Boettcher (loc.\
cit., cf.\ also some older studies \cite{BG} or newer developments
\cite{Mali}) merely replaced ${\cal T}$ by its product with parity
${\cal P}$ and conjectured that the above-mentioned and
physically well-motivated weakening of Hermiticity could be most
appropriately characterized as an antilinear `symmetry' or
`$\cal PT$-symmetry', ${\cal PT}H = H{\cal PT}$, of all the
Hamiltonians in question. Equivalently \cite{205} one may speak
about the $\cal P$-pseudo-Hermiticity defined by the relation
\be
  H^\dagger
  ={\cal P}\,H\,{\cal P}^{-1}\,.
  \label{eq:PTS}
\ee
\par
%
%---------------------------------------------------
%
In this paper we intend to concentrate on
implementing the resulting $\cal PT$-symmetric SUSYQM
factorization scheme in the case of the `simplest' model which remains
`realistic' and `solvable' at the same time. This means that our
`initial'  Schr\"{o}dinger equation
\be
  \left[- \frac{d^2}{d x^2}+V^r(x)+{\rm i}V^i(x)\right]\psi(x)
  =E\psi(x)\label{eq:problem}
\ee
(where we dropped the superscript `$(+)$' as temporarily
redundant) will contain just the most trivial infinitely deep
square-well form
\be
  V^r(x)=\left\{\begin{array}{ll}
      +\infty & x<-L\\[0.1cm]
      0 & -L<x<L\\[0.1cm]
      +\infty& x>L\end{array}\right. \label{eq:V-r}
\ee
for the real part of the potential and the most elementary short-range
one
\be
  V^i(x)=\left\{\begin{array}{ll}
       0 & x<-l\\[0.1cm]
       -g & -l<x<0\\[0.1cm]
       +g& 0<x<l,\\[0.1cm]
       0& x>l
       \end{array}\right.
       \qquad l<L , \qquad  g>0\,
       \label{eq:V-i}
\ee
for its imaginary part. As a consequence of (\ref{eq:V-r}), the
wavefunctions will be defined on a finite interval $(-L, L)$ with a
variable length $2L$, on which they satisfy the standard Dirichlet
boundary conditions \cite{ptsqw,Langer}
\be
  \psi(\pm L)=0 .\label{eq:dirichlet}
\ee
\par
%
%----------------------------------------------------------
%
Given the background of the result obtained in \cite{twopoint}, we
derive in section \ref{sec:ptsw}, an elegant trigonometric form of
the standard matching conditions for wavefunctions at the
discontinuities of the potential (subsection \ref{sec:secular})
and discuss the practical semi-numerical determination of the
energies with arbitrary precision (subsection
\ref{sec:graphical}).\par
%
%--------------------------------------------------------------------
%
In section \ref{sec:SUSYQM} we address the key concern of our
present paper, viz., the investigation of the problem in the
context of SUSYQM. Here the non-Hermiticity and discontinuities
create some specific features, which are dealt with in detail.
After deriving the superpotential and the partner potential in
subsection \ref{sec:W}, we construct the eigenfunctions of the
latter and analyze the discontinuities in subsections
\ref{sec:eigenfunctions} and \ref{sec:discontinuities},
respectively.\par
%
%--------------------------------------------------------------------
%
Some physical aspects of our results are finally discussed in
more detail in section~\ref{sec:discussion}.\par
%
%========================================
%
\section{\boldmath Trigonometric secular equation
\label{sec:ptsw}}
\setcounter{equation}{0}

\subsection{\boldmath $\cal PT$-symmetric square well inside a real one
\label{sec:secular}}

Let us denote the four regions $-L < x < -l$, $-l < x < 0$, $0 < x
< l$, $l < x < L$ by $L2$, $L1$, $R1$, $R2$, respectively. We
shall henceforth append these symbols as subscripts to all
quantities pertaining to such regions. The complex potential
$V(x)$, defined in equations (\ref{eq:V-r}) and (\ref{eq:V-i}),
may therefore be rewritten as
\be
  V_{L2}(x) = 0 \qquad V_{L1}(x) = - {\rm i} g \qquad V_{R1}(x) = {\rm i} g \qquad
  V_{R2}(x) = 0. \label{eq:potential}
\ee
\par
%
%-----------------------------------------------------------------
%
The general solution of~(\ref{eq:problem}) satisfying the
conditions (\ref{eq:dirichlet}) can be written as
\be
  \psi(x) = \left\{\begin{array}{l}
    \psi_{L2}(x) = A_L \sin[k(L+x)] \\[0.1cm]
    \psi_{L1}(x) = B_L \cosh(\kappa^* x) + {\rm i} \frac{C_L}{\kappa^* l} \sinh(\kappa^*
         x) \\[0.1cm]
    \psi_{R1}(x) = B_R \cosh(\kappa x) + {\rm i} \frac{C_R}{\kappa l} \sinh(\kappa x)
         \\[0.1cm]
    \psi_{R2}(x) = A_R \sin[k(L-x)]
  \end{array} \right. \label{eq:psi}
\ee
where
\be
  \kappa=s+{\rm i}t \qquad E=k^2=t^2-s^2 \qquad g=2st. \label{eq:k-kappa}
\ee
Here $s,\ t$ and $k$ are real and, for the sake of definiteness,
are assumed positive. A priori, $A_L$, $B_L$, $C_L$, $A_R$, $B_R$
and $C_R$ are some complex constants.\par
%
%---------------------------------------------------
%
On assuming that $\cal PT$-symmetry is unbroken, we obtain the conditions
\be
  \psi^*_{L2}(-x) = \psi_{R2}(x) \qquad \psi^*_{L1}(-x) = \psi_{R1}(x) \label{eq:psi-PT}
\ee
from which we get
\be
  A_L^* = A_R \equiv A \qquad B_L^* = B_R \equiv B \qquad C_L^* = C_R \equiv C.
  \label{eq:ABC}
\ee
The derivative of (\ref{eq:psi}), taking (\ref{eq:ABC}) into
account, reads
\be
  \partial_x \psi(x)=\left\{\begin{array}{l}
    \partial_x \psi_{L2}(x) = k A^* \cos[k(L+x)] \\[0.1cm]
    \partial_x \psi_{L1}(x) = \kappa^* B^* \sinh(\kappa^*x) + {\rm i} \frac{C^*}{l} \cosh
        (\kappa^* x) \\[0.1cm]
    \partial_x \psi_{R1}(x) = \kappa B \sinh(\kappa x) + {\rm i} \frac{C}{l}\cosh(\kappa x)
        \\[0.1cm]
    \partial_x \psi_{R2}(x) = - k A \cos[k(L-x)]
  \end{array}\right.. \label{eq:derivative}
\ee
\par
%
%-----------------------------------------------------------------------
%
Let us now match the wavefunction and its derivative at $x=0$ and
impose $\cal PT$-symmetry in the neighbourhood of the origin:
\be
  \psi_{R1}(0) = \psi_{L1}(0) \in \R \qquad \partial_x \psi_{R1}(0) = \partial_x \psi_{L1}(0)
  \in {\rm i} \R.
\ee
This leads to
\be
  B, C \in \R.  \label{eq:cond-BC}
\ee
\par
%
%-----------------------------------------------------------------------
%
It now remains to match $\psi$ and $\partial_x \psi$ at $x=\pm l$.
Since $\psi$ is $\cal PT$-symmetric, it is enough to impose
matching conditions at $x=l$:
\be
  \psi_{R2}(l) = \psi_{R1}(l) \qquad \partial_x \psi_{R2}(l) =  \partial_x \psi_{R1}(l).
\ee
This yields
\bea
  A\sin [k(L-l)] & = & B \cosh(\kappa l) + {\rm i} \frac{C}{\kappa l} \sinh(\kappa l)
       \label{eq:A-BC1} \\[0.1cm]
  -k A \cos[k(L-l)] & = & \kappa B \sinh(\kappa l) + {\rm i} \frac{C}{l} \cosh(\kappa l).
       \label{eq:A-BC2}
\eea
\par
%
%---------------------------------------------------------------------
%
We conclude that the final form of $\psi$ is
\be
  \psi(x) = \left\{\begin{array}{l}
    \psi_{L2}(x) = A^* \sin[k(L+x)] \\[0.1cm]
    \psi_{L1}(x) = B \cosh(\kappa^* x) + {\rm i} \frac{C}{\kappa^* l} \sinh(\kappa^*
         x) \\[0.1cm]
    \psi_{R1}(x) = B \cosh(\kappa x) + {\rm i} \frac{C}{\kappa l} \sinh(\kappa x)
         \\[0.1cm]
    \psi_{R2}(x) = A \sin[k(L-x)]
  \end{array} \right. \label{eq:psi-bis}
\ee
where the complex constant $A$ is determined by one of the
equations (\ref{eq:A-BC1}) and (\ref{eq:A-BC2}), while the real
constants $B$ and $C$ have to satisfy a condition obtained by
eliminating $A$ between (\ref{eq:A-BC1}) and (\ref{eq:A-BC2}):
\bea
  && \kappa l B \{k\cos[k(L-l)] \cosh(\kappa l)  + \kappa \sin[k(L-l)] \sinh(\kappa l)\}
       \nonumber \\
  && \mbox{} + {\rm i} C \{k \cos[k(L-l)] \sinh(\kappa l) + \kappa \sin[k(L-l)]
       \cosh(\kappa l)\} = 0. \label{eq:rel-BC}
\eea
We may therefore express both constants $A$ and $C$ in terms of
$B$ as
\bea
  A & = & B\, \frac{\kappa \csc[k(L-l)] \csch(\kappa l)}{k \cot[k(L-l)] + \kappa
         \coth(\kappa l)} \label{eq:A} \\
  C & = & {\rm i} \kappa l B\, \frac{k \cot[k(L-l)] \coth(\kappa l) + \kappa}
         {k \cot[k(L-l)] + \kappa \coth(\kappa l)}. \label{eq:C}
\eea
\par
%
%--------------------------------------------------------------------------
%
Since, from (\ref{eq:cond-BC}), the left-hand side of equation
(\ref{eq:C}) is real, the same should be true for the right-hand
one. The resulting condition can be written as
\bea
  && k^2 \cot^2[k(L-l)] [\kappa \coth(\kappa l) + \kappa^* \coth(\kappa^* l)]
      \nonumber \\
  && \mbox{} + k \cot[k(L-l)] [\kappa^2 + 2 \kappa \kappa^* \coth(\kappa l)
      \coth(\kappa^* l) + \kappa^{*2}] \nonumber \\
  && \mbox{} + \kappa \kappa^* [\kappa \coth(\kappa^* l) + \kappa^* \coth(\kappa
      l)] = 0.  \label{eq:C-real}
\eea
On expressing $k^2$, $\kappa$ and $\kappa^*$ in terms of $s$ and
$t$ through equation (\ref{eq:k-kappa}) and using some elementary
trigonometric identities, condition (\ref{eq:C-real}) is easily
transformed into
\bea
  && k \sin[2k(L-l)] [s^2 \cosh(2sl) + t^2 \cos(2tl)] \nonumber \\
  && \mbox{} - \cos[2k(L-l)] [s^3 \sinh(2sl) - t^3 \sin(2tl)] \nonumber \\
  && \mbox{} + s t^2 \sinh(2sl) - s^2 t \sin(2tl) = 0 \label{eq:transcendental}
\eea
where $k = \sqrt{t^2 - s^2}$.
\par
%
%+++++++++++++++++++++++++++++++++++++++++++++++++++++++++
%
\subsection{Graphical and numerical determination of the energies
\label{sec:graphical}}
%
%----------------------------------------------------------------------
%
The transcendental equation (\ref{eq:transcendental}) has to be
complemented by the constraint  (\ref{eq:k-kappa}),
\be
  s t = \frac{1}{2} g .\label{eq:hyperbola}
\ee
The couples of roots $(s_n, t_n)$, $n=0$, 1, 2,~\ldots, of this
pair of equations define all the bound-state energies $E_n$ by the
elementary formula
\be
  E_n = t_n^2 - s_n^2 \qquad n=0, 1, 2, \ldots.  \label{eq:E}
\ee
In practice, the $(s_n, t_n)$ values may be obtained as the
intersection points in the $(s, t)$ plane of the curves
representing the roots of the transcendental equation
(\ref{eq:transcendental}) with the hyperbola
(\ref{eq:hyperbola}).\par
%
%--------------------------------------------------------------------------
%
Before proceeding to discuss the graphical and numerical
determination of $E_n$ in general, it is worth reviewing three
interesting limiting cases of equation (\ref{eq:transcendental}).
%
%--------------------------------------------------------------
%
The first one corresponds to the limit $l \to L$, wherein the
present square well with three matching points reduces to the one
with a single discontinuity. Equation (\ref{eq:transcendental})
then simply becomes
\be
  s \sinh(2sL) + t \sin(2tL) = 0
\ee
which coincides with equation (9) of \cite{ptsqw} (where $g$ is
denoted by $Z$ and $L=1$).\par
%
%-----------------------------------------------------------
%
The second limiting case corresponds to $l \to 0$ and gives back
the real square well. Since the constraint (\ref{eq:hyperbola})
then disappears, we are only left with equation
(\ref{eq:transcendental}) acquiring the simple form
\be
  \sin(2kL) = 0.
\ee
Its solutions are provided by the hyperbolas $t^2 - s^2 =
\left(\frac{n\pi}{2L}\right)^2$, $n=1$, 2,~\ldots, where the $n=0$
value is discarded because no acceptable wavefunction can be
associated with it. We therefore arrive at the well-known
quadratic spectrum $E_n^2 = \left(\frac{n\pi}{2L}\right)^2$,
$n=1$, 2,~\ldots, of the real square well.\par
%
%-----------------------------------------------------------
%
The existence of the third special limiting regime is connected
with the bounded nature of our imaginary barrier (\ref{eq:V-i}).
In the language of perturbation theory this means \cite{Langer}
that the influence of this barrier on the values of the energies
(\ref{eq:E}) weakens quickly with the growth of the quantum number
$n$. At the higher excitations, as a consequence, the
$n-$dependence of the energies will not deviate too much from the
$l \to 0$ rule $E_n \sim n^2 \gg 1$. In the other words, the
growth of $n$ will imply the growth of $t_n \sim n \gg 1$ and the
decrease and smallness of the roots $s_n = g/(2t_n) \ll 1$. In
this regime, we may imagine that $k = t\,\sqrt{1 - s^2/t^2}= t -
s^2/(2t) + {\cal O}(s^4/t^3)= t-g^2/(8t^3) + {\cal O}(1/n^7)$ so
that the six components of our quantization condition
(\ref{eq:transcendental}), {\it viz.},
\bea
  &&  s^2 k \sin[2k(L-l)]\cosh(2sl)
  + t^2 k \sin[2k(L-l)] \cos(2tl) \nonumber \\
  && \mbox{} - s^3 \cos[2k(L-l)] \sinh(2sl)
+  t^3 \cos[2k(L-l)]
    \sin(2tl) \nonumber \\
  && \mbox{} + s t^2 \sinh(2sl) - s^2 t \sin(2tl) = 0
\nonumber
  \label{eq:appranscend}
\eea
may be characterized by their asymptotic sizes ${\cal O}(1/n) $,
${\cal O}(n^3) $, ${\cal O}(1/n^4) $, ${\cal O}(n^3) $, ${\cal
O}(n^0) $ and ${\cal O}(1/n) $, respectively. Once we omit all the
negligible ${\cal O}(1/n) $ terms and insert $s = g/(2t)$ whenever
necessary, we arrive at the thoroughly simplified approximate
secular equation
\be
      \sin(2kL) +
      \frac{g^2l}{2k^3}
      +
       {\cal O}\left (\frac{1}{k^4}\right )= 0 .\label{eq:apprcend}
\ee
Its roots are easily found,
\be
      k=k_n=\frac{\pi\,n}{2L} + (-1)^{n+1}
      \frac{2g^2lL^2}{\pi^3 n^3}
      + {\cal O}\left (\frac{1}{n^4}\right ),
      \label{eq:aend}
\ee
and give
\be
      E_n=k_n^2=\left (\frac{\pi\,n}{2L}\right )^2 + (-1)^{n+1}
      \frac{2g^2lL}{\pi^2 n^2}
      + {\cal O}\left (\frac{1}{n^3}\right )
      \label{eq:nd}
\ee
i.e., a nice and elementary approximate energy formula for all the
highly excited states.

In the general case, the bound-state energies (\ref{eq:E}) of our
model are determined from the simultaneous solutions of equations
(\ref{eq:transcendental}) and (\ref{eq:hyperbola}). Although the
former is transcendental, one of its roots is quite obvious,
namely $s=t$. When we realize that this implies $k=0$ and
substitute the solution into equations (\ref{eq:psi-bis}) --
(\ref{eq:C}), we obtain a vanishing wavefunction. This is in
accordance with an insight provided by the Hermitian limit
$g\rightarrow 0$ or $l \to 0$.\par
%
%------------------------------------------------------------------------------------
%
%%\newsavebox{\figa}
  %%  \savebox{\figa}{
%%   \rotatebox{0}{\scalebox{0.9}{
  %%  \includegraphics*{figa.eps}
%%    }}
  %%  }
%% \newsavebox{\pfiga}
   %% \savebox{\pfiga}{
%% \noindent\begin{tabular}{p{60mm}}
%
 %   \small{Figure 1a: Solutions of~(\ref{eq17}) form the semi-ovals.
  %  Their intersections with the hyperbola $2st=g$ determine energy levels
%% $E=k^2=t^2-s^2$ of the system. Here $g=$, $l=0.$.  }\\
%
 %%   \end{tabular}}
%
The other solutions of~(\ref{eq:transcendental}) can be found
numerically and graphically. As we can see in figure 1 where we
work with re-scaled length units in which $L=1$, they form
semi-ovals in $(s,t)$ plane. We can observe the absence of
robustly real energy levels, i.e., levels remaining real for any
value of $g$, which played their role in~\cite{twopoint}.\par
%
%-----------------------------------------------------------------------------------
%
The locally decreasing character of the semi-oval maxima could
cause a complexification of higher energy pairs while the lower
pairs would remain real. In other words, the semi-oval maxima
might be decreasing faster then the hyperbola
(\ref{eq:hyperbola}). This race in decrease can be judged easily
when we use a hyperbolic coordinate system. As shown in figure 2,
in this setting, the maxima prove to increase monotonically while
the hyperbola is represented by a horizontal straight line.
%
%%\newsavebox{\figb}
  %%  \savebox{\figb}{
%%    {\scalebox{0}{
  %%  \includegraphics*{figb.eps}
%%    }}
  %%  }
%%\newsavebox{\pfigb}
  %%  \savebox{\pfigb}{
%%\noindent\begin{tabular}{p{60mm}}
%
  %%  \small{Figure 1b: The previous picture(Fig.1a) in $[ts,\sqrt{t^2-s^2}]$ plane.
    %% }\\
%
 %%   \end{tabular}}
%
%%\begin{center}\hspace{-10mm}\begin{tabular}[h]{ll}
  %%                  \usebox{\figa}&\usebox{\figb}\\
    %%                \usebox{\pfiga}&\usebox{\pfigb}\\
      %%              \end{tabular}\end{center}
%
%------------------------------------------------------------------
%
Consequently, our model preserves a sequential merging of the
energy levels. The critical value $g_c$ of the coupling constant
$g$, for which the two lowest energy levels merge together, is of
high importance. It is the boundary of exact $\cal PT$-symmetry,
which we consider to be physically relevant and assumed in
deriving equation~(\ref{eq:transcendental}). For a higher value of
$g$, the wavefunction $\cal PT$-symmetry would be broken.\par
%
%--------------------------------------------------------------------
%
We found $g_c$ for various values of the parameter $l$. Since
$g_c$ rises rapidly as $l\rightarrow 0$, we present its values in
combination of graph and table (see figure 3 and table 1).
%
%%\newsavebox{\figc}
  %%  \savebox{\figc}{
%%    \rotatebox{0}{\scalebox{1}{
  %%  \includegraphics*{figc.eps}
%%    }}
  %%  }
%
%%\newsavebox{\pfigc}
  %%  \savebox{\pfigc}{
%%\noindent\begin{tabular}{p{80mm}}
%
  %%  \small{Figure 2: Fifty values of critical coupling $g_c$. As $l\rightarrow 0$, $g_c$
%%   increases rapidly. }\\
%
  %%  \end{tabular}}
%%\newsavebox{\ptab}
  %%  \savebox{\ptab}{
%%\noindent\begin{tabular}{p{140mm}}
%
  %%  \small{Table 1: Numerical values of critical coupling $g_c$ independence
    %%on parameter $l$. The values suggest that the coupling rises faster then $1/l$ for
%%$l\rightarrow 0$.
  %%  The value of $g_c$ for $l=1$ coincides with~\cite{gezawell}.}\\
%
    %%\end{tabular}}
%
%
%%\begin{center}
%%\begin{tabular}{c}
  %%                  \usebox{\figc}\\
    %%                \usebox{\pfigc}\\
      %%              \end{tabular}\end{center}
%
%%\begin{center}\begin{tabular}{c}
%%\begin{tabular}{|c||c|c|c|c|c|c|c|c|c|}\hline
  %%  $l$&1.00 &0.70 &0.50 &0.40 &0.30 &0.20 &0.10 &0.01&0.001 \\
%%    \hline
  %%  $g_c\sim$&4.4753 &4.8129 &6.4364 &8.6011 &13.426 &27.273 &95.832
%% &9895.4&486950 \\
  %%  \hline
%%\end{tabular}\\
%%\usebox{\ptab}\\
%%\end{tabular}
%% \end{center}
%
As the parameter $l$ approaches zero, $g_c$ tends to infinity and
the semi-oval maxima run to infinity as well. As explained in
subsection \ref{sec:secular}, equation (\ref{eq:transcendental})
then provides the bound-state energies of the real square well. On
the other hand, for $l \to L=1$, we get back the critical coupling
$g_c \simeq 4.4753$, previously obtained for the square well
in~\cite{ptsqw} and \cite{gezawell}.\par
%
%============================================
%
\section{\boldmath The SUSY partner potential
\label{sec:SUSYQM}} \setcounter{equation}{0}

The purpose of the present section is to construct and study the
SUSY partner  $H^{(-)}$ of the square-well Hamiltonian $H^{(+)}$,
defined in equation (\ref{eq:potential}), in the
physically-relevant unbroken $\cal PT$-symmetry regime,
corresponding to $g < g_c$.
%
%+++++++++++++++++++++++++++++++++++++++
%
\subsection{Determination of the parameters \label{sec:W}}

Identifying $V^{(+)}$ with the square-well potential
(\ref{eq:potential}), i.e., $V^{(+)}_{L2}(x) = 0$,
$V^{(+)}_{L1}(x) = - {\rm i} g$, $V^{(+)}_{R1}(x) = {\rm i} g$,
$V^{(+)}_{R2}(x) = 0$ and $E_0 = k_0^2 = t_0^2 - s_0^2 = -
\kappa_0^2 + {\rm i} g$, we obtain for the superpotential and the
partner potential the results
\be
  W(x) = \left\{\begin{array}{l}
      W_{L2}(x) = k_0 \tan[k_0(x + x_{L2})] \\[0.1cm]
      W_{L1}(x) = - \kappa_0^* \tanh[\kappa_0^*(x + x_{L1})] \\[0.1cm]
      W_{R1}(x) = - \kappa_0 \tanh[\kappa_0(x - x_{R1})] \\[0.1cm]
      W_{R2}(x) = k_0 \tan[k_0(x - x_{R2})]
  \end{array} \right.
\ee
and
\be
  V^{(-)}(x) = \left\{\begin{array}{l}
      V^{(-)}_{L2}(x) = 2 k_0^2 \sec^2[k_0(x + x_{L2})] \\[0.1cm]
      V^{(-)}_{L1}(x) = - 2 \kappa_0^{*2} \sech^2[\kappa_0^*(x + x_{L1})] - {\rm i} g
           \\[0.1cm]
      V^{(-)}_{R1}(x) = - 2 \kappa_0^2 \sech^2[\kappa_0(x - x_{R1})] + {\rm i} g  \\[0.1cm]
      V^{(-)}_{R2}(x) = 2 k_0^2 \sec^2[k_0(x - x_{R2})]
  \end{array} \right.
  \label{eq:partner-0}
\ee
respectively. Here $x_{L2}$, $x_{L1}$, $x_{R1}$ and $x_{R2}$
denote four integration constants.\par
%
%----------------------------------------------------------------------
%
We now choose $x_{L2}$ and $x_{R2}$ as
\be
  x_{L2} = L + \frac{\pi}{2k_0} \qquad x_{R2} = L - \frac{\pi}{2k_0} \label{eq:integration-2}
\ee
to ensure that $V^{(-)}_{L2}$ and $V^{(-)}_{R2}$ blow up at the
end points $x=-L$ and $x=L$. This is in tune with~\cite{CQ}. We
thus get
\be
  V^{(-)}_{L2}(x) = 2 k_0^2 \csc^2[k_0(x + L)] \qquad V^{(-)}_{R2}(x) = 2 k_0^2
  \csc^2[k_0(x - L)]. \label{eq:partner-bis}
\ee
Observe that for the superpotential, $W_{L2}(x)$ and $W_{R2}(x)$
also blow up at these points:
\be
  W_{L2}(x) = - k_0 \cot[k_0(x + L)] \qquad W_{R2}(x) = - k_0 \cot[k_0(x - L)].
\ee
\par
%
%-------------------------------------------------------------------------
%
Let us next consider the unbroken SUSY condition
(\ref{eq:unbroken-susy}), where according to (\ref{eq:psi-bis})
the ground-state wavefunction of $H^{(+)}$ is given by
\bea
  \psi^{(+)}_{0R2}(x) & = & \psi^{(+)*}_{0L2}(-x) = A^{(+)}_0 \sin[k_0(L - x)] \\
  \psi^{(+)}_{0R1}(x) & = & \psi^{(+)*}_{0L1}(-x) = B^{(+)}_0 \cosh(\kappa_0 x) + {\rm i}
       \frac{C^{(+)}_0}{\kappa_0 l} \sinh(\kappa_0 x).
\eea
Note that the superscript `$(+)$' is appended to the wavefunction
and the coefficients to signify that we are dealing with
Hamiltonian $H^{(+)}$. It is straightforward to see that equation
(\ref{eq:unbroken-susy}) is automatically satisfied in the regions
$R2$ and $L2$ due to the choice made for the integration constants
$x_{R2}$, $x_{L2}$ in equation (\ref{eq:integration-2}). On the
other hand, in the region $R1$ we find a condition fixing the
value of $x_{R1}$,
\be
  \tanh(\kappa_0 x_{R1}) = - \frac{{\rm i} C^{(+)}_0}{\kappa_0 l B^{(+)}_0} = \frac{k_0
  \cot[k_0(L-l)] \coth(\kappa_0 l) + \kappa_0}{k_0 \cot[k_0(L-l)] + \kappa_0
  \coth(\kappa_0 l)} \label{eq:x_R1}
\ee
where in the last step we used equation (\ref{eq:C}). A similar
relation applies in $L1$, thus leading to the result
\be
  x_{L1} = x_{R1}^*. \label{eq:x_L1}
\ee
\par
%
%--------------------------------------------------------------------------------
%
Note that in contrast with the real integration constants
$x_{R2}$, $x_{L2}$, the constants $x_{R1}$ and $x_{L1}$ are
complex. Separating both sides of equation (\ref{eq:x_R1}) into a
real and an imaginary part, we obtain the two equations
\bea
  \frac{\sinh X \cosh X}{\cosh^2 X \cos^2 Y + \sinh^2 X \sin^2 Y} & = & \frac{N^r}{D}
       \label{eq:x_R1-1} \\
  \frac{\sin Y \cos Y}{\cosh^2 X \cos^2 Y + \sinh^2 X \sin^2 Y} & = & \frac{N^i}{D}
       \label{eq:x_R1-2}
\eea
where we have used the decompositions $\kappa_0 = s_0 + {\rm
i}t_0$, $x_{R1} = x_{R1}^r + {\rm i} x_{R1}^i$, $\kappa_0 x_{R1} =
X + {\rm i} Y$, implying that
\be
  X = s_0 x_{R1}^r - t_0 x_{R1}^i \qquad Y = t_0 x_{R1}^r + s_0 x_{R1}^i
\ee
and we have defined
\be
  N^r  = \{- s_0^2 \cos[2k_0(L-l)] + t_0^2\} \sinh(2s_0 l) + k_0 s_0 \sin[2k_0(L-l)]
       \cosh(2s_0 l)
\ee
\be
  N^i  = \{s_0^2 - t_0^2  \cos[2k_0(L-l)]\} \sin(2t_0 l) - k_0 t_0 \sin[2k_0(L-l)]
       \cos(2t_0 l)
\ee
\bea
  D & = & \{- s_0^2 \cos[2k_0(L-l)] + t_0^2\} \cosh(2s_0 l) + \{s_0^2 - t_0^2
        \cos[2k_0(L-l)]\} \cos(2t_0 l) \nonumber \\
  && \mbox{} + k_0 \sin[2k_0(L-l)] [s_0 \sinh(2s_0 l) + t_0 \sin(2t_0 l)].
\eea
Equations (\ref{eq:x_R1-1}) and (\ref{eq:x_R1-2}), when solved
numerically, furnish the values of both the parameters $x_{R1}^r$
and $x_{R1}^i$.\par
%
%-------------------------------------------------
%
One may also observe that the resulting superpotential $W(-x) = - W^*(x)$
and partner potential $V^{(-)}(-x) = V^{(-)*}(x)$ are $\cal
PT$-antisymmetric and $\cal PT$-symmetric, respectively.\par
%
%++++++++++++++++++++++++++++++++++++++++++++
%
\subsection{Eigenfunctions in the partner potential \label{sec:eigenfunctions}}

On exploiting the first intertwining relation in
(\ref{eq:intertwining}), the eigenfunctions $\psi^{(-)}_n(x)$,
$n=0$, 1, 2,~\ldots, of $H^{(-)}$ can be obtained by acting with
$\ca$ on $\psi^{(+)}_{n+1}(x)$, subject to the preservation of the
boundary and continuity conditions
\bea
  \psi^{(-)}_{nL2}(-L) & = & 0 \qquad \psi^{(-)}_{nR2}(L) = 0 \label{eq:boundary} \\
  \psi^{(-)}_{nL2}(-l) & = & \psi^{(-)}_{nL1}(-l) \qquad \partial_x \psi^{(-)}_{nL2}(-l) =
        \partial_x \psi^{(-)}_{nL1}(-l) \label{eq:continuity-1} \\
  \psi^{(-)}_{nL1}(0) & = & \psi^{(-)}_{nR1}(0) \qquad \partial_x \psi^{(-)}_{nL1}(0) =
        \partial_x \psi^{(-)}_{nR1}(0) \label{eq:continuity-2} \\
  \psi^{(-)}_{nR1}(l) & = & \psi^{(-)}_{nR2}(l) \qquad \partial_x \psi^{(-)}_{nR1}(l) =
        \partial_x \psi^{(-)}_{nR2}(l). \label{eq:continuity-3}
\eea
%
%---------------------------------------------------------------------------
%
Application of $\ca$ leads to the forms
\bea
  \psi^{(-)}_{nL2}(x) & = & C^{(-)}_{nL2}\, A^{(+)*}_{n+1} \sin[k_{n+1}(L+x)]\nonumber \\
  && \mbox{} \times \{k_{n+1} \cot[k_{n+1}(L+x)] - k_0 \cot[k_0(L+x)]\}
        \label{eq:partner-psi-1} \\
  \psi^{(-)}_{nL1}(x) & = & C^{(-)}_{nL1}\, B^{(+)}_{n+1} \sinh(\kappa_{n+1}^* x)
        \{\kappa_{n+1}^* - \kappa_0^* \tanh[\kappa_0^*(x + x_{R1}^*)]
        \coth(\kappa_{n+1}^* x)\} \nonumber \\
  && \mbox{} + C^{(-)}_{nL1}\, \frac{{\rm i} C^{(+)}_{n+1}}{\kappa_{n+1}^* l}
        \sinh(\kappa_{n+1}^* x) \nonumber \\
  && \mbox{} \times \{\kappa_{n+1}^* \coth(\kappa_{n+1}^* x) - \kappa_0^*
        \tanh[\kappa_0^*(x + x_{R1}^*)]\} \\
  \psi^{(-)}_{nR1}(x) & = & C^{(-)}_{nR1}\, B^{(+)}_{n+1} \sinh(\kappa_{n+1} x)
        \{\kappa_{n+1} - \kappa_0 \tanh[\kappa_0(x - x_{R1})]
        \coth(\kappa_{n+1} x)\} \nonumber \\
  && \mbox{} + C^{(-)}_{nR1}\, \frac{{\rm i} C^{(+)}_{n+1}}{\kappa_{n+1} l}
        \sinh(\kappa_{n+1} x) \nonumber \\
  && \mbox{} \times \{\kappa_{n+1} \coth(\kappa_{n+1} x) - \kappa_0
        \tanh[\kappa_0(x - x_{R1})]\} \\
  \psi^{(-)}_{nR2}(x) & = & C^{(-)}_{nR2}\, A^{(+)}_{n+1} \sin[k_{n+1}(L-x)]\nonumber \\
  && \mbox{} \times \{- k_{n+1} \cot[k_{n+1}(L-x)] + k_0 \cot[k_0(L-x)]\}
        \label{eq:partner-psi-4}
\eea
where $C^{(-)}_{nL2}$, $C^{(-)}_{nL1}$, $C^{(-)}_{nR1}$,
$C^{(-)}_{nR2}$ denote some complex constants and equation
(\ref{eq:x_L1}) has been used. It can be easily checked that the
boundary conditions (\ref{eq:boundary}) are automatically
satisfied by these eigenfunctions. It therefore remains to impose
the continuity conditions (\ref{eq:continuity-1}) --
(\ref{eq:continuity-3}).\par
%
%-----------------------------------------------------------------------------
%
Let us first match the regions $L1$ and $R1$ at $x=0$. The
continuity conditions (\ref{eq:continuity-2}) yield the two
relations
\be
  C^{(-)}_{nR1} \left[B^{(+)}_{n+1} \kappa_0 \tanh(\kappa_0 x_{R1}) + \frac{{\rm i}
  C^{(+)}_{n+1}}{l}\right] = C^{(-)}_{nL1} \left[- B^{(+)}_{n+1} \kappa_0^* \tanh(
  \kappa_0^* x_{R1}^*) + \frac{{\rm i} C^{(+)}_{n+1}}{l}\right] \label{eq:LR-1}
\ee
\bea
  && C^{(-)}_{nR1} \left\{B^{(+)}_{n+1} [\kappa_{n+1}^2 - \kappa_0^2 \sech^2(\kappa_0
        x_{R1})] + \frac{{\rm i} C^{(+)}_{n+1}}{l} \kappa_0 \tanh(\kappa_0 x_{R1})\right\}
        \nonumber \\
  && = C^{(-)}_{nL1} \left\{B^{(+)}_{n+1} [\kappa_{n+1}^{*2} - \kappa_0^{*2}
        \sech^2(\kappa_0^* x_{R1}^*)] - \frac{{\rm i} C^{(+)}_{n+1}}{l} \kappa_0^*
        \tanh(\kappa_0^* x_{R1}^*)\right\}. \label{eq:LR-2}
\eea
Since equations (\ref{eq:x_R1}) and (\ref{eq:k-kappa}) provide the
two constraints
\bea
  \kappa_0 \tanh(\kappa_0 x_{R1}) & = & - \kappa_0^* \tanh(\kappa_0^* x_{R1}^*) \\
  \kappa_{n+1}^{*2} - \kappa_{n+1}^2 & = & \kappa_0^{*2} - \kappa_0^2 = - 2g
\eea
equations (\ref{eq:LR-1}) and (\ref{eq:LR-2}) are compatible and
lead to the condition
\be
  C^{(-)}_{nR1} = C^{(-)}_{nL1}.
\ee
\par
%
%---------------------------------------------------------------------------
%
Considering next the matching between $R1$ and $R2$ at $x=l$, we
obtain from equation (\ref{eq:continuity-3}) the two conditions
\bea
  && C^{(-)}_{nR1} \{k_{n+1} \cot[k_{n+1}(L-l)] + \kappa_0 \tanh[\kappa_0(l -
        x_{R1})]\} \nonumber \\
  && = C^{(-)}_{nR2} \{k_{n+1} \cot[k_{n+1}(L-l)] - k_0 \cot[k_0(L-l)]\} \label{eq:R12-1}
\eea
\bea
  && C^{(-)}_{nR1} \biggl(\kappa_{n+1}^2 - \kappa_0^2 + \kappa_0 \tanh[\kappa_0(l -
       x_{R1})] \{k_{n+1} \cot[k_{n+1}(L-l)] \nonumber \\
  && \quad\mbox{} + \kappa_0 \tanh[\kappa_0(l - x_{R1})]\}\biggr) \nonumber \\
  && = C^{(-)}_{nR2} \biggl(k_0^2 - k_{n+1}^2 - k_0 \cot[k_0(L-l)] \{k_{n+1}
       \cot[k_{n+1}(L-l)] \nonumber \\
  && \quad\mbox{} - k_0 \cot[k_0(L-l)]\}\biggr) \label{eq:R12-2}
\eea
after making use of equations (\ref{eq:A}) and (\ref{eq:C}) to
eliminate $A^{(+)}_{n+1}$, $B^{(+)}_{n+1}$ and $C^{(+)}_{n+1}$.
Equations (\ref{eq:R12-1}) and (\ref{eq:R12-2}) both yield the
same result
\be
  C^{(-)}_{nR1} = C^{(-)}_{nR2} \label{eq:C-R12}
\ee
due to the two relations
\be
  \kappa_0 \tanh[\kappa_0(l - x_{R1})] = - k_0 \cot[k_0(L-l)] \label{eq:relation-1}
\ee
and
\be
  \kappa_{n+1}^2 - \kappa_0^2 = k_0^2 - k_{n+1}^2
\ee
deriving from (\ref{eq:x_R1}) and (\ref{eq:k-kappa}),
respectively.\par
%
%--------------------------------------------------------------------
%
Since a result similar to (\ref{eq:C-R12}) applies at the
interface between regions $L2$ and $L1$, we conclude that the
partner potential eigenfunctions are given by equations
(\ref{eq:partner-psi-1}) -- (\ref{eq:partner-psi-4}) with
\be
  C^{(-)}_{nL2} = C^{(-)}_{nL1} = C^{(-)}_{nR1} = C^{(-)}_{nR2} \equiv C^{(-)}_n.
\ee
Such eigenfunctions are $\cal PT$-symmetric provided we choose
$C^{(-)}_n$ imaginary:
\be
  C^{(-)*}_n = - C^{(-)}_n.
\ee
\par
%
%++++++++++++++++++++++++++++++++++++++++++
%
\subsection{Discontinuities in the partner potential \label{sec:discontinuities}}

In subsection \ref{sec:W}, we have constructed the SUSY partner
$V^{(-)}(x)$ of a piece-wise potential with three discontinuities
at $x = -l$, 0 and $l$. We may now ask the following question:
does the former have the same discontinuities as the latter or
could the discontinuity number decrease? We plan to prove here
that the second alternative can be ruled out.\par
%
%-----------------------------------------------------------------------
%
{}For such a purpose, we will examine successively under which
conditions $V^{(-)}(x)$ could be continuous at $x=l$ or at $x=0$
and we will show that such restrictions would not be compatible
with some relations deriving from the unbroken-SUSY assumption
(\ref{eq:unbroken-susy}). Observe that we do not have to study
continuity at $x = -l$ separately, since $V^{(-)}(x)$ being $\cal
PT$-symmetric must be simultaneously continuous or discontinuous
at $x = -l$ and $x=l$.\par
%
%---------------------------------------------------------------------
%
Let us start with the point $x=l$. Matching there
$V^{(-)}_{R1}(x)$ and $V^{(-)}_{R2}(x)$, given in equations
(\ref{eq:partner-0}) and (\ref{eq:partner-bis}), respectively,
leads to the relation
\be
  - 2 \kappa_0^2 \sech^2[\kappa_0 (l - x_{R1})] + {\rm i} g = 2 k_0^2 \csc^2[k_0(L-l)].
\ee
On using (\ref{eq:relation-1}) and some simple trigonometric
identities, such a relation can be transformed into $k_0^2 = -
\kappa_0^2 + \frac{1}{2} {\rm i} g$, which manifestly contradicts
equation (\ref{eq:k-kappa}). Hence continuity of $V^{(-)}(x)$ at
$x=l$ is ruled out.\par
%
%-----------------------------------------------------------------------
%
Consider next the point $x=0$. On equating $V^{(-)}_{R1}(0)$ with
$V^{(-)}_{L1}(0)$ and employing (\ref{eq:partner-0}) and
(\ref{eq:x_L1}), we obtain the condition
\be
  - 2 \kappa_0^2 \sech^2(\kappa_0 x_{R1}) + {\rm i} g = - 2 \kappa_0^{*2}
  \sech^2(\kappa_0^* x_{R1}^*) - {\rm i} g.
\ee
Equations (\ref{eq:cond-BC}) and (\ref{eq:x_R1}) then yield the
relation $- \kappa_0^2 + \frac{1}{2} {\rm i} g = - \kappa_0^{*2} -
\frac{1}{2} {\rm i} g$, which contradicts equation
(\ref{eq:k-kappa}) again. Continuity of $V^{(-)}(x)$ at $x=0$ is
therefore excluded too.\par
%
%----------------------------------------------------------------------------
%
We conclude that under the simplest assumption of unbroken SUSY
with a factorization energy equal to the ground-state energy of
$H^{(+)}$, the partner potential $V^{(-)}(x)$ has the same three
discontinuities at $x = -l$, 0 and $l$ as $V^{(+)}(x)$.\par
%
%===========================================
%
\section{Discussion \label{sec:discussion}}
\setcounter{equation}{0}

Among all the $\cal PT$-symmetric models, field-theoretical
background explains the lasting interest in the purely imaginary
long-range model $V(x)= {\rm i} x^3$ \cite{Bessis,DDT} and its
generalizations $V(x)=x^2({\rm i} x)^\delta$ with the imaginary
part $V^i(x)$ exhibiting, at any $\delta \in [0,2)$, a
characteristic `strongly non-Hermitian' (SNH) long-range growth in
`coordinate' $x \in \R$. Up to the harmonic oscillator at
$\delta=0$, all of the latter SNH $\cal PT$-symmetric models are
only solvable by approximate methods. Still, rigorous proofs exist
showing that their spectra are all real \cite{DDT}.\par
%
%-----------------------------------------------------------------
%
By rigorous means, the reality of the spectrum has also been shown
for many other $\cal PT$-symmetric potentials $V$. Some of them
turn out to be exactly solvable \cite{SI,BR,BQ},  and those for
which $V^i(\pm \infty)=0$ may be called 'weakly non-Hermitian'
(WNH). Their WNH character is reflected not only by a less
explicit influence of the imaginary part of the potential upon the
spectrum, but also by the existence of SUSY partners \cite{BMQ1,
BR, BMQ2} which, in some special cases, may be real and Hermitian
\cite{Andrianov,BR}.\par
%
%-----------------------------------------------------------------
%
In the light of similar observation one might feel tempted to
perceive WNH models as `partially compatible' with our intuitive
expectations. This impression may be further enhanced by noticing
that another exactly solvable model, viz., the typical WNH spiked
form of the $\delta=0$ harmonic oscillator, as described in
\cite{ptho}, proved of particular interest in the SUSYQM context
as well~\cite{crea,BMQ2}.\par
%
%-----------------------------------------------
%
Potentials $V(x)$ with shapes that are piece-wise constant may be
considered equally exceptional. All of these square-well-type
models with forces located inside a finite interval $(-L,L)$ may
be easily classified by the number of their discontinuities.\par
%
%------------------------------------------------
%
The simplest nontrivial non-Hermitian square-well potential must
have at least one discontinuity (= matching point at $x=0$). While
the real part of this $V$ is just a trivial shift of the energy
scale, it may be kept equal to zero. Then, the non-zero strength
$Z$ of the spatially antisymmetric and purely imaginary $V$ is the
only free (real) parameter of the whole model with SNH features
\cite{CQ,ptsqw}. Its $\cal PT$-symmetry remains unbroken in an
interval of $Z \in (-Z_{crit}, Z_{crit})$ while its ground-state
energy becomes complex beyond $Z_{crit} \approx 4.48$ (in standard
units $\hbar = 2m = 1$~\cite{ptsqw,gezawell}).\par
%
%-------------------------------------------------------------------
%
It is known that some of these features are generic \cite{Langer}.
Quantitatively, their occurrence has also been confirmed for the
twice-constant SNH model $V$ with two discontinuities
\cite{twopoint}. Qualitatively, all of these observations
facilitate the applicability and physical interpretation of the
piece-wise constant models significantly \cite{Batal}, especially
because the numerical values of the maximal allowed couplings
prove to be, in general, quite large. This allows us to guarantee
the (necessary) reality of the energies by keeping simply our
choice of $Z$ safely below this maximum. \par
%
%----------------------------------------------------
%
The family of WNH square-well models may only start at the
piece-wise potential with three discontinuities. In our present
study of such a model it was important to demonstrate the
parallelism of its properties with the exact solutions of the {\em
smooth} complex potentials of similar shapes \cite{crea}.\par
%
%--------------------------------------------------------
%
The most obvious parallel lies in the observation that a key
formal feature of the SUSY partners $H^{(\pm)}$ is that they may
remain both non-Hermitian and $\cal PT$-symmetric. Of course, the
parity ${\cal P}$ cannot define the positive-definite norm
\cite{Mali,Langer,srni,Bpriv}. A consistent physical
interpretation of the similar non-Hermitian models was recently
agreed (cf., e.g., \cite{BBJ}) to lie in the existence of {\em a
new} metric-like operator ${\cal P}_{(+)}>0$ which is positive
definite. This Hermitian operator may be assumed to play the role
of the `physical' metric \cite{Geyer}. This means that once our
equation (\ref{eq:PTS}) is satisfied by the old Hamiltonian and by
the new, {\em positive-definite} metric ${\cal P}_{(+)}$, we may
declare the underlying quantum Hamiltonian quasi-Hermitian,
leading to the standard probabilistic interpretation of the theory
(cf.\ the recent discussions of some related subtleties in
\cite{Kretschmer}). Against this background our attention has been
concentrated upon the feasibility of bound-state construction in a
model with a phenomenologically appealing shape of the
potential.\par
%
%------------------------------------------------------------------
%
A couple of consequences may be expected. Our model may open the
way towards addressing one of the most difficult problems
encountered in $\cal PT$-symmetric quantum mechanics \cite{Bpriv},
viz., the control of a possible instability of the spectrum
reality \cite{twopoint,fragile}. Indeed, due to the
pseudo-Hermiticity property (\ref{eq:PTS}) of our Hamiltonians
$H$, the energies need not be real (i.e., observable) in principle
\cite{205}.\par
%
%----------------------------------------------------------------
%
Our WNH model may be also characterized by the simplicity of the
bound-state wavefunctions. This allowed us to construct the
superpotential yielding access,  rather easily, to the Witten-type
SUSY hierarchy. In this regard the compact form of our
trigonometric secular equation was welcome and particularly
important, especially for any future projects trying to connect
the mathematical $\cal PT$-symmetry with physical
phenomenology.\par
%
%---------------------------------------------
%
In such a perspective, the most challenging {mathematical}
problems attached to the non-Hermitian models descend from the
reality of their exceptional points \cite{Heiss}. The simplest
solvable models of the square-well type seem to offer a
transparent laboratory for their study since  the indeterminate
auxiliary pseudo-metric $\cal P$ coincides with the common
parity.\par
%
%-----------------------------------------------
%
In the context of physics, the phenomenological appeal of all the
piece-wise constant analogues of the purely imaginary cubic force
represented a strong motivation for the systematic constructions
of the positive-definite metric operators of \cite{Geyer}
(cf.\ also \cite{Mali,205,Batal}). In particular,  the highly
appealing factorized form  ${\cal P}_{(+)}= {\cal CP}> 0$ of these
metric operators has been used and, for physical reasons, the
factor $\cal C$ itself has been called `charge' (cf.\ \cite{BBJ}). For all
the models with relevance in field theory (like $V \sim ix^3$), the
constructions of $\cal C$ were shown feasible by WKB and perturbative
methods~\cite{joness}.\par
%
%----------------------------------------------------------------------
%
In comparison, the solvability of all the simpler models
facilitates the construction of $\cal C$ (called, usually,
quasi-parity in this context \cite{SI,ptho,srni,Quesne}). An
interesting energy-shift interpretation of the quasi-parity (which
is a new symmetry of the Hamiltonian) emerged in the strongly
spiked short-range model considered in~\cite{Omar}.\par
%
%----------------------------------------------------------------------
%
After we return to the square-well models, the quasi-parity or
charge operator $\cal C$ may be constructed in the specific form
which differs sufficiently significantly from the unit operator
just in a finite-dimensional subspace of the Hilbert space
\cite{Langer,twopoint,Batal}. This is one of the most important
merits of this class of models. It seems to open a new inspiration
for a direct physical applicability of non-Hermitian models
whenever their spectrum remains real. \par
%
%======================================
%
\subsection*{Acknowledgements}
The participation of HB, VJ and MZ complied with the Institutional
Research Plan AV0Z10480505. CQ is a Research Director, National
Fund for Scientific Research (FNRS), Belgium. VJ was supported by
the project no. 2388G-6 of FRVS. MZ was supported by the grant
A1048302 of GA AS.\par
%
%=============================================
%

\newpage

\section*{Figure  captions}
\par

\vspace{1.2cm}

\subsection*{Figure 1: Solutions of~(\ref{eq:transcendental}) form the semi-ovals.
    Their intersections with the hyperbola $2st=g$ determine energy levels
 $E=k^2=t^2-s^2$ of the system. Here $g=650$ and $l=0.04$.
   }

\subsection*{Figure 2: The previous picture (Fig.1) in $[ts,k]$ plane, where {$ k=\sqrt{t^2-s^2} $}.
     We set $g=650$ and $l=0.04$ again.}

\subsection*{Figure 3: Fifty values of critical couplings $g_c$,
increasing rapidly as $l$ decreases, $l\rightarrow 0$. }

\par

\vspace{2cm}

\section*{Table captions}

\par

\vspace{1.2cm}

\subsection*{Table 1: Numerical values of $g_c$ in dependence
    on the parameter $l$. The table suggests that the critical coupling
    grows
    faster than $1/l$ for small
$l$.
   }

\par

\vspace{2cm}

\section*{Table 1}

\begin{center}\begin{tabular}{c}
\begin{tabular}{|c||c|c|c|c|c|c|c|c|c|}\hline
    $l$&1.00 &0.70 &0.50 &0.40 &0.30 &0.20 &0.10 &0.01&0.001 \\
    \hline
    $g_c\sim$&4.4753 &4.8129 &6.4364 &8.6011 &13.426 &27.273 &95.832
 &9895.4&486950 \\
    \hline
\end{tabular}\\
\end{tabular}
 \end{center}
\par
%
%==================================================================
%

\end{document}